\documentstyle[12pt]{article}
\renewcommand{\baselinestretch}{1.2}
\begin{document}

%	==========
%	 Abstract
%	==========

%\pagebreak
\pagenumbering{arabic}
\setcounter{page}{1}
\centerline{\large \bf Influence of Implementation on the Properties of}
\centerline{\large \bf Pseudorandom Number Generators}
\centerline{\large \bf with a Carry Bit}

\vskip1cm
\centerline{I. Vattulainen$^{1,3}$, K. Kankaala$^{1,2}$,
            J. Saarinen$^1$, and T. Ala-Nissila$^{1,3}$}
\bigskip

\centerline{\em $^1$Department of Electrical Engineering}
\centerline{\em Tampere University of Technology}
\centerline{\em P.O. Box 692}
\centerline{\em FIN - 33101 Tampere, Finland}

\bigskip

\centerline{\em $^2$Centre for Scientific Computing}
\centerline{\em P.O. Box 405, FIN - 02100 Espoo, Finland}

\bigskip

\centerline{\em $^3$Research Institute for Theoretical Physics}
\centerline{\em P.O. Box 9 (Siltavuorenpenger 20 C)}
\centerline{\em FIN - 00014 University of Helsinki, Finland}

\begin{abstract}
We present results of extensive statistical and bit level tests
on three implementations of a pseudorandom number generator algorithm using
the lagged Fibonacci method with an occasional addition of an extra bit.
First implementation is the RCARRY generator of James,
which uses subtraction. The second is a modified version of it,
where a suggested error present in the original
implementation has been corrected.
The third is our modification of RCARRY such that
it utilizes addition of the carry bit.
Our results show that there are no significant differences
between the performance of these three generators.
\end{abstract}


\pagebreak
\pagenumbering{arabic}
\setcounter{page}{1}
\parskip=0.3cm
\renewcommand{\baselinestretch}{1.2}

%       ==============
%	 Introduction
%       ==============

\section{Introduction}

Random numbers are needed in various applications,
including cryptography \cite{Dav89}, stochastic optimization \cite{Aar89}, and
Monte Carlo methods \cite{Bin92}. Because of practical reasons
random numbers are usually produced by deterministic rules,
implemented as pseudorandom number generators. In spite
of their fully deterministic origin the quality of
pseudorandom numbers may often be good enough for many
applications.

To confirm the suitability of a given pseudorandom number generator
for practical use, it should be subjected to a rigourous test program
which reveals the strengths and weaknesses of the algorithm and,
in particular, its {\it implementation}. Recently, such an extensive
test program has been carried out by the present authors \cite{Vat93b}.
By performing a comparative evaluation using statistical, bit level
and visual tests we were able to assess the quality of a group of
random number generators which are commonly used in the applications
of physics.

One of the generators included in Ref. \cite{Vat93b} was RCARRY,
which uses the so called {\em ``subtract-and-borrow''} algorithm
which has been implemented by James \cite{Jam90}. In the tests,
RCARRY clearly displayed the poorest statistical properties of
the generators tested, suggesting possible problems in the
implementation. Supporting this, James has recently
reported \cite{Jam93} the observation of M. L\"uscher that the
original implementation of RCARRY may contain a small error, which
may adversely affect the quality of the random number sequence.
The purpose of the present work is to address this issue. To this end,
we present results of extensive statistical and bit level tests on
the corrected version of RCARRY, and compare the results to those
of Ref. \cite{Vat93b}. In addition, we test a slightly different version
of the RCARRY generator, which uses an {\it ``add-and-carry''} algorithm
based on the {\it addition} of a carry bit. We call this generator
ADCARRY. Our results reveal that there is very little
difference between the statistical properties of the original RCARRY
and its corrected version, as well as the ADCARRY generator.
All these generators display a relatively poor performance in two
of the gap tests presented here.

\section{Implementation of the Generators}

The three pseudorandom number generators tested in this work
are based on a lagged Fibonacci algorithm, which is augmented
by an occasional addition of a carry bit. The basic formula is:
\begin{equation}
X_{i} = (X_{i-24}\ \pm\ X_{i-10}\ \pm c) \mbox{ mod } b.
\end{equation}
The carry bit $c$ is zero if the sum is less than or equal to $b$,
and otherwise ``$c=1$ in the least significant bit position''
\cite{Jam90}. The choice for $b$ is $2^{24}$.
The period of the generator is about $2^{1407}$ \cite{Jam90}
and it produces random numbers distributed between [0,1).
Only the 24 most significant bits are guaranteed to be good.

The inclusion of the carry bit $c$ in the lagged Fibonacci algorithm
was done in order to improve its properties \cite{Mar90b}.
Recently, however, it has been shown \cite{Cou,Tez} that this
type of algorithms are in fact equivalent to linear congruential
generators with very large prime moduli. Consequently, they
inherit unfavourable lattice structures in higher dimensions.

The original implementation of Eq. (1) was done by James \cite{Jam90},
based on the ideas of Marsaglia {\em et al.} \cite{Mar90b}. It uses
the subtraction contained in Eq. (1). In this work, we shall denote
it by I1. The second generator I2 includes the suggested
correction of L\"uscher and James, who recommend replacing
line 13 of the code of Ref. \cite{Jam90}

\begin{tabular}{p{2cm} l}
 & {\tt uni = seeds(i24) - seeds(j24) - carry,}
\end{tabular}

by

\begin{tabular}{p{2cm} l}
 & {\tt uni = seeds(j24) - seeds(i24) - carry}.
\end{tabular}

The third generator ADCARRY (I3) uses the operation known as
``add-and-carry'', in which subtraction in Eq. (1) has been replaced by
addition. In this version the lines 13 - 15 of \cite{Jam90} are
rewritten as:

\begin{tabular}{p{2cm} l}
 & {\tt uni = seeds(j24) + seeds(i24) + carry}\\
 & {\tt if(uni.ge.1.) then}\\
 & \mbox{\hspace{2cm}} {\tt uni = uni - 1.}
\end{tabular}

Otherwise, the implementation is identical to that of RCARRY
\cite{Lusch1,Lusch2}.

\section{Test methods}

Tests scrutinizing the quality of random numbers can be divided into
three main categories: statistical tests \cite{Knu81}, bit level tests
\cite{Alt88,Mar85,Vat93b}
for testing the properties of random numbers on binary level, and
visual tests \cite{Knu81} which may give some further qualitative
information on the statistical properties of random numbers.
A number of these tests were implemented and employed extensively in Ref.
\cite{Vat93b}. In this work, we have
repeated the same statistical tests for I2 and I3. They
are listed in Table 1, where the numbering refers to the
parameters of Ref. \cite{Vat93b}. From bit level tests, only the
$d$-tuple test \cite{Alt88,Mar85} was done since it was shown to be
sufficient. Finally, the random numbers were plotted in two
dimensions for purposes of visual inspection.

The test bench is described in detail in Ref. \cite{Vat93b}.
Description of the statistical tests can also be found in Ref. \cite{Knu81}.
In brief, the statistical accuracy of all the tests
was improved by utilizing a one way Kolmogorov - Smirnov test
\cite{Knu81} to a large number (1000 or more) of test statistics.
This approach has been realized earlier by L'Ecuyer \cite{Lec88}.
The final test variables are therefore the values $K^+$ and $K^-$ of
a Kolmogorov - Smirnov test statistic $K$ \cite{Knu81}. In each test
the generator was considered to fail the test if the observed descriptive
level $\delta = P(K \leq \{ K^+, K^-\} | H_0)$ was less than
0.05 or larger than 0.95.

\section{Results}

Results of the statistical tests for the descriptive levels
$\delta^+$ and $\delta^-$ are summarized in Table 2, where
the numbering refers to Table 1. In each test the chosen generator was
initialized with the seed 667790. In case a failure occurred, the
generator was subjected to another test starting from the final
state of the first test. If another immediate failure occurred,
the generator was tested for the third time starting from a new
state with an initial seed 14159 (from the decimals of $\pi$).

In Table 2, frames with thin lines indicate a single failure,
frames with double single lines two consequtive failures, and frames with
bold lines three consequtive failures in the corresponding tests.
The results of the original RCARRY ({\cal I1}) by James \cite{Jam90}
are shown on the left (from Ref. \cite{Vat93b}), whereas the results
of the corrected version ({\cal I2}) and ADCARRY ({\cal I3})
are at the center and on the right, respectively.

Based on the results, it is clear that the corrected version
of RCARRY (I2) using
arithmetic subtraction performs no significantly better than the
original RCARRY (I1). The main malady of RCARRY,
namely the clear failing of the gap tests 6 and 8 with parameters
$\alpha = 0, \beta = 0.05$ and $\alpha = 0.95, \beta = 1$ \cite{Knu81},
respectively, is still characteristic of I2. This signifies
the existence of local correlations in the vicinity of zero and one.
The same conclusion applies to ADCARRY as well, signaling basic
problems with these algorithms.

In the $d$-tuple test each implementation was tested two times and
the bits considered failed had two consequtive failures. The results
are shown in Table 3. In our notation, bit number one is the
most significant bit (excluding the sign bit). For the original
implementation of RCARRY (I1) only the 24 most significant bits
are guaranteed to be good, which the tests confirm \cite{Vat93b}.
The implementation I2 yields identical results, whereas
ADCARRY ({\cal I3}) gives only 22 good bits (see Fig. 3).

Finally, visual tests on bit level support the results above. In Figs.
1, 2 and 3 we show subsequent random numbers for I1, I2 and I3
in binary form on a $120 \times 120$ matrix, when only 24 most significant
bits are included. No clear correlations are visible,
except for the last two bits of ADCARRY where strong correlations are
apparent. No visual indications of the suggested \cite{Cou,Tez}
lattice structure were found in these generators.

\section{Summary and conclusions}

In this work, we have compared the results of detailed
statistical and bit level tests for three implementations of
random number algorithms using a lagged Fibonacci sum with the
addition of a carry bit. Results for RCARRY and its corrected
version show very little difference.
Also, a new generator ADCARRY
using purely additive arithmetics fares no better statistically, and
has two good bits less than RCARRY. Fortunately enough,
these bits are not among the most significant ones.
Overall, our results suggest
that the basic algorithm of Eq. (1) on which these generators are based
seems to lead to observable correlations. The persistent failure of this
class of generators in the gap tests may lead to problems in some
applications, e.g. in lattice simulations \cite{Luscher}.

Finally, we would like to emphasize the importance of extensive
testing such as presented here before using {\it any} new
pseudorandom number generator. Even a good algorithm can be corrupted
by a poor implementation, as we have previously demonstrated \cite{Vat93b}.
Hence, a good amount of
scepticism towards pseudorandom number generators without extensive
test results seems prudent. It should also be noted that even when
no statistical or bit level correlations are found, direct physical
tests of random number generators should be used to reveal possible
``hidden'' correlations \cite{Vat93c,Fer92}.

\clearpage

{\Large {\bf Acknowledgments}}

\medskip

We would like to thank Fred James and Martin L\"uscher
for fruitful correspondence.
The Centre for Scientific Computing Ltd., Tampere University
of Technology, and University of Helsinki
have provided ample computing resources. This research
has been supported by the Academy of Finland.

E-mail addresses:
{\tt
Ilpo.Vattulainen@csc.fi,
Kari.Kankaala@csc.fi,\\
jukkas@ee.tut.fi,
ala@phcu.helsinki.fi}

\pagebreak

\clearpage

{\Large {\bf Table captions}}

\medskip

{\sc Table 1}.
List of the statistical tests. Numbers refer to the choice of parameters
in Ref. \cite{Vat93b}.

{\sc Table 2}.
Results of the statistical tests. I1, I2 and I3 refer to
RCARRY, its corrected version, and ADCARRY, respectively.
Results for RCARRY are from Ref. \cite{Vat93b}.
Depicted numbers are the observed descriptive levels $\delta^+$
and $\delta^-$ of the
test variables $K^+$ and $K^-$, respectively.
A generator was considered to fail the test if
the descriptive level was less than 0.05 or more than 0.95. Single,
double and triple consequtive
failures are indicated by single, double, and bold lines,
respectively. The numbers shown are from the first run only.

{\sc Table 3}.
Results of the bit level $d$-tuple test.
The bits marked failed have failed the test twice.
See text for details.

\bigskip\bigskip

{\Large {\bf Figure captions}}

\medskip

{\sc Figure 1}.
24 bit binary representation of random numbers produced by
implementation RCARRY (I1) on a $120 \times 120$ matrix.

{\sc Figure 2}.
24 bit binary representation of random numbers produced by
implementation I2 of RCARRY on a $120 \times 120$ matrix.

{\sc Figure 3}.
24 bit binary representation of random numbers produced by
ADCARRY on a $120 \times 120$ matrix.

\clearpage

{\hfill Table I \hfill }

\begin{table}[h]
\normalsize \centering
\begin{tabular}{| c | l |}
\hline
Test 	& Test method 	\\
number 	&		\\
\hline\hline
1	& Equidistribution test (1) \\
2	& Equidistribution test (2) \\ \hline
3	& Serial test in 2 dimensions \\
4	& Serial test in 3 dimensions \\
5	& Serial test in 4 dimensions \\ \hline
6	& Gap test (1) \\
7	& Gap test (2) \\
8	& Gap test (3) \\ \hline
9	& Maximum of $t$  test, $t=5$\\
10	& Maximum of $t$  test, $t=3$\\ \hline
11	& Collision test (1) \\
12	& Collision test (2) \\
13	& Collision test (3) \\ \hline
14	& Runs-up test \\ \hline
\end{tabular}
\end{table}

\clearpage

\clearpage

\smallskip

{\hfill Table III \hfill }

\begin{table}[h]
\normalsize \centering
\begin{tabular}{| c | c | c |}
\hline
Implementation	& Failing	& Number of \\
		& bits		& ``good'' bits  \\
\hline\hline
I1	& $25-31$ & 24 \\ \hline
I2	& $25-31$ & 24 \\ \hline
I3	& $23-31$ & 22 \\ \hline
\end{tabular}
\end{table}


\begin{thebibliography}{250}
\bibitem{Lusch1}
In a private communication, M. L\"uscher has pointed out that the
period for ADCARRY is unknown and
it is difficult to prove analytically that it is always large.
\end{thebibliography}
\end{document}